# RADIO  BRIDGES  OF  THE  FUTURE BERWEEN SOLAR SYSTEM AND  THE  NEAREST  50  STARS

**Claudio  Maccone**

Director for Scientific Space Exploration, International Academy of Astronautics (IAA)
Associate, Istituto Nazionale di Astrofisica (INAF, Italy)
E-mails**:** **claudio.maccone@gmail.com** and **clmaccon@libero.it**

**Nicolò  Antonietti**

Corresponding Member of the International Academy of Astronautics (IAA)
Associate, Istituto Nazionale di Astrofisica (INAF, Italy)
E-mail: **nicolo.antonietti@gmail.com**

**ABSTRACT.**  The Solar Gravitational Lens (SGL) is a gift of nature that Humanity is now ready to exploit.

Though SGL physics started with Einstein's 1936 paper on the gravitational lensing provided by every star, it was not until 1979 that the idea of a space mission reaching the Sun's nearest focal sphere at 550 Astronomical Units (AU) was put forward by Von Eshleman. By the year 2000, the senior author of this paper (CM) had submitted a relevant formal proposal to ESA about thee relevant space mission to 550 AU. He presented his ideas at NASA-JPL for the first time on August 18, 1999: **(32) Claudio Maccone - The Sun's Gravity lens and its use for Interstellar Exploration - YouTube** and by 2009 he had published his book **Deep Space Flight and Communications - Exploiting the Sun as a Gravitational Lens | Claudio Maccone | Springer**, that was translated into Chinese by 2014 and finally awarded the IAA Book Award in 2018 **(32) Claudio Maccone - Speech at IAA Dinner in Paris on March 26 2019 - YouTube**. In 2020 NASA awarded a $2million grant to JPL to prepare for the first FOCAL space mission.

But radio bridges between the Sun and any nearby star may also be conceived: **(32) Claudio Maccone: Breakthrough Discuss 2016 – FOCAL Missions to 550 AU Insuring Interstellar LINKS - YouTube**. The idea is that, if Humanity will be able to send unmanned space probes to the nearest stars in the future, each of these probes could be placed behind the star of arrival and along the star-Sun line, thus allowing for TWO gravitational lenses to work together. That will result in a permanent communication system with much REDUCED POWERS to keep the radio link between the two stellar systems: a veritable Galactic Internet.

In this paper, we study for the first time the 100 radio bridges between the Sun and each of the nearest 100 stars in the Galaxy. Of course, this work is for the centuries to come. But knowing which natural radio bridge between the Sun and each of the nearest 100 stars is MORE CONVENIENT, will open the ROAD MAP for the HUMAN EXPANSION into the Galaxy.

## 1.  A SHORT INTRODUCTION TO THE SOLAR GRAVITATIONAL LENS (SGL)

In a 2010 paper (ref. [1]), this author pointed out that, before shooting for even the nearest stars, Humanity will have to reach the "modest" distances between 550 and 1000 AU to take full advantage of the Sun as a gravitational lens. The huge magnifying power of the Sun's lens on all electromagnetic waves, in fact, would allow us to get in advance very detailed radio and optical pictures of any stellar system to be later reached by the actual interstellar missions. See ref. [1] for more details.

During the 2009 International Astronautical Congress (IAC) held at Daejeon, Republic of Korea, October 12th-16th, 2009, this author also presented one more paper, ref. [2], enlightening the "communication counterpart" of ref. [1]. In other words, the Sun's gravity lens was then turned to the full advantage of interstellar radio communications. These prove to be feasible only by resorting to both the gravity lenses of the Sun and of the target star system and so we refer to this possibility as to the "Interstellar Radio Bridges" between the Sun and any nearby star, like Alpha Cen A, Bernard's Star, Sirius A, etc. These topics will be explored in the coming sections of this paper.

As short historical introduction seems, however, desirable:

1) The gravitational focusing effect of any star according to general relativity was first described by Albert Einstein in 1936, but his work was virtually forgotten until 1964, when his ideas were applied to the focusing by a galaxy located between the Earth and a distant cosmological object, like a quasar.

2) Subsequent astronomical discoveries of actual gravitational lenses since 1978 eliminated all doubts about the focusing effect as predicted by general relativity.

3) In 1979 this theory was first applied to the case of the Sun as the lensing object. Thus, the possibility was realized of sending a spacecraft to the minimal focal distance of 550 AU to exploit the enormous magnifications upon electromagnetic waves of all kinds (both radio and optical) provided by the gravitational lens of the Sun, as shown in Figure 1.

4) The relevant space mission was dubbed FOCAL (an acronym for "Fast Outgoing Cyclopean Astronomical Lens") by the author and proposed to space agencies. It is studied in detail in his 2009, treatise "Deep Space Flight and Communications – Exploiting the Sun as a Gravitational Lens" (ref. [6]). This book embodies all previous material published about the FOCAL space mission and updates it. The book was translated into Chinese by 2014 and was awarded the IAA Book Award in 2018, see the relevant ceremony held in Paris on March 26th, 2019: **(32) Claudio Maccone - Speech at IAA Dinner in Paris on March 26 2019 - YouTube**

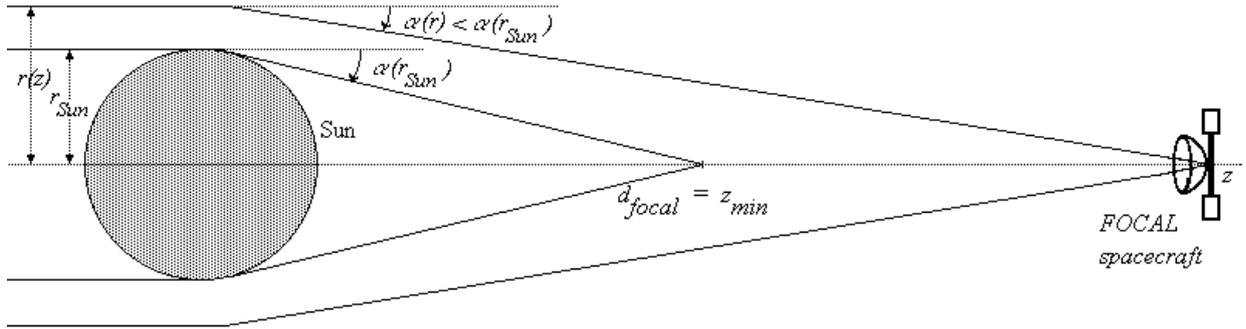

**Figure 1. Geometry of the Sun gravitational lens with the minimal focal length of 550 AU (= 3.17 light days = 13.75 times beyond Pluto's orbit) and the FOCAL spacecraft position beyond the minimal focal length.**

5) From the Schwarzschild solution to the Einstein equations of general relativity, it can be proven that the minimal focal distance, $d_{focal}$, that the FOCAL probe must reach, is given by

$$d_{focal} \approx \frac{c^2}{4G} \cdot \frac{r_{Sun}^2}{M_{Sun}} . \qquad (1)$$

This basic result may also be rewritten in terms the *Schwarzschild radius*

$$r_{Schwarzschild} = \frac{2GM_{Sun}}{c^2}, \qquad (2)$$

yielding

$$d_{focal} \approx \frac{r_{Sun}^2}{2\, r_{Schwarzschild}} . \qquad (3)$$

Numerically, one finds

$$d_{focal} \approx \approx 550 \text{ AU} \approx 3.17 \text{ light days} . \qquad (4)$$

*These are the fundamental formulae yielding the minimal focal distance of the gravitational lens of the Sun, i.e. the minimal distance from the Sun's center that the FOCAL spacecraft must reach in order to get magnified radio pictures of whatever lies on the other side of the Sun with respect to the spacecraft position.*

A simple, but important consequence of the above discussion is that *all points on the straight line beyond*

*this minimal focal distance are foci too*, because the light rays passing by the Sun further than the minimum distance have smaller deflection angles and thus come together at an even greater distance from the Sun. And the very important astronautical consequence of this fact for the FOCAL mission is that *it is not necessary to stop the spacecraft at 550 AU. It can go on to almost any distance beyond and focus as well or better.* In fact, the further it goes beyond 550 AU the less distorted are the radio waves crossing the Sun Corona fluctuations. Chapter 6 of ref. [6] may be seen for many astrophysical details about the Sun Corona, related effects, and astronautical consequences for the FOCAL space mission.

## 2. THE HUGE (ANTENNA) GAIN OF THE GRAVITATIONAL LENS OF THE SUN

Having thus determined the minimal distance of 550 AU that the FOCAL spacecraft must reach, one now wonders what's the good of going so far out of the solar system, i.e. how much focussing of light rays is caused by the gravitational field of the Sun. The answer to such a question is provided by the technical notion of "antenna gain", that stems out of antenna theory.

A standard formula in antenna theory relates the antenna gain, $G_{antenna}$, to the antenna effective area, $A_{effective}$, and to the wavelength $\lambda$ or the frequency $\nu$ by virtue of the equation (refer, for instance, to ref. [7], in particular page **6**-117, equation (6-241)):

$$G_{antenna} = \frac{4\pi A_{effective}}{\lambda^2}. \quad (5)$$

Now, assume the antenna is circular with radius $r_{antenna}$, and assume also a 50% efficiency. Then, the antenna effective area is obviously given by

$$A_{effective} = \frac{A_{physical}}{2} = \frac{\pi r_{antenna}^2}{2}. \quad (6)$$

Substituting this back into (5) yields the antenna gain as a function of the antenna radius and of the observed frequency:

$$G_{antenna} = \frac{4\pi A_{effective}}{\lambda^2} = \frac{2\pi A_{physical}}{\lambda^2} = \frac{2\pi^2 r_{antenna}^2}{\lambda^2} = \frac{2\pi^2 r_{antenna}^2}{c^2} \cdot \nu^2. \quad (7)$$

The important point here is that *the antenna gain increases with the square of the frequency*, thus favoring observations on frequencies as high as possible.

Is anything similar happening for the Sun's gravitational lens also? *Yes* is the answer, and the "gain" (one maintains this terminology for convenience) of the gravitational lens of the Sun can be proved to be

$$G_{Sun} = 4\pi^2 \frac{r_{Schwarzschild}}{\lambda} \quad (8)$$

or, invoking the expression (2) of the Schwarzschild radius

$$G_{Sun} = \frac{8\pi^2 G M_{Sun}}{c^2} \cdot \frac{1}{\lambda} = \frac{8\pi^2 G M_{Sun}}{c^3} \cdot \nu. \quad (9)$$

The mathematical proof of equation (8) is difficult to achieve. The author, unsatisfied with the treatment of this key topic given in ref. [8], turned to three engineers of the engineering school in his home town, Renato Orta, Patrizia Savi and Riccardo Tascone. To his surprise, in a few weeks they provided a full proof of not just the Sun gain formula (8), but also of the focal distance for rays originated from a source at finite distance (not mentioned in this paper, see equation (1.10), page 7 of ref. [8]). Their proof is fully described in ref. [8], and is based on the aperture method used to study the propagation of electromagnetic waves, rather than on ray optics. Using the words of these three authors' own Abstract, they have "computed the radiation pattern of the [spacecraft] Antenna+Sun system, which has an extremely high directivity. It has been observed that the focal region of the lens for an incoming plane wave is a half line parallel to the propagation direction starting at a point [550 AU] whose position is related to the blocking effect of the Sun disk (Figure 1). Moreover, a characteristic of this thin lens is that its gain, defined as the magnification factor of the antenna gain, is constant along this half line. In particular, for a wavelength of 21 cm, this lens gain reaches the value of 57.5 dB. Also a measure of the transversal extent of the focal region has been obtained. The performance of this radiation system has been determined by adopting a thin lens model which introduces a phase factor depending on the logarithm of the impact parameter of the incident rays. Then the antenna is considered to be in transmission mode and the radiated field is computed by asymptotic evaluation of the radiation integral in the Fresnel approximation".

One is now able to compute the Total Gain of the Antenna+Sun system, that is simply obtained by multiplying equations the two equations yielding the spacecraft gain proportional to $\nu^3$ and the Sun gain proportional to $\nu$:

$$G_{Total} = G_{Sun} \cdot G_{antenna} = \frac{16\pi^4 G M_{Sun} r_{antenna}^2}{c^5} \cdot \nu^3$$
(10)

Since the total gain increases with the *cube* of the observed frequency, it favors electromagnetic radiation in the microwave region of the spectrum. The table in Figure 2 shows the numerical data provided by the last equation for five selected frequencies: the hydrogen line at 1420 MHz and the four frequencies that the Quasat radio astronomy satellite planned to observe, had it been built jointly by ESA and NASA as planned before 1988, but Quasat was abandoned by 1990 due to lack of funding. The definition of dB is of course:

$$N \, dB = 10 \log_{10} N = 10 \ln N / \ln 10.$$

| Line | Neutral Hydrogen | | OH radical | | $H_2O$ |
|---|---|---|---|---|---|
| **Frequency** $\nu$ | **1420 MHz** | **327 MHz** | **1.6 GHz** | **5 GHz** | **22 GHz** |
| Wavelength $\lambda$ | 21 cm | 92 cm | 18 cm | 6 cm | 1.35 cm |
| S/C Antenna Beamwidth | 1.231 deg | 5.348 deg | 1.092 deg | 0.350 deg | 0.080 deg |
| Sun Gain | 57.4 dB | 51.0 dB | 57.9 dB | 62.9 dB | 69.3 dB |
| 12-meter Antenna S/C Gain | 42.0 dB | 29.3 dB | 43.1 dB | 53.0 dB | 65.8 dB |
| **Combined Sun + S/C Gain** | **99.5 dB** | **80.3 dB** | **101.0 dB** | **115.9 dB** | **135.1 dB** |

**Figure 2: Table showing the gain of the Sun's lens alone, the gain of a 12-meter spacecraft (S/C) antenna and the combined gain of the Sun+S/C Antenna system the at five selected frequencies important in radioastronomy.**

## 1. THE RADIO LINK

The goal of this paper is to prove that only by exploiting the Sun as a gravitational lens will we be able to have reliable telecommunication links across large interstellar distances. In other words, a direct link between different star systems, even if held by virtue of the largest radiotelescopes on Earth, will not be feasible across distances on the order of thousand of light years or more. We want to show that only a FOCAL mission in the direction from the Sun opposite to that target star system will ensure a reliable telecommunication link across thousands of light years. Namely, we prove that Bit Error Rate (or BER, see site http://www.en.wikipedia.org/wiki/Bit_error_rate) will be unacceptable already at the distance of Alpha Centauri unless we resort to a supporting FOCAL space mission in the opposite direction from the Sun.

In order to face these problems mathematically, we must first understand the radio link among any two stars, and we think that no neater treatment of this subject exists than the book "Radio Astronomy" by the late professor John D. Kraus of Ohio State University (ref. [7]), that we follow hereafter.

Consider a radio transmitter that radiates a Power $P_t$ isotropically and uniformly over a bandwidth $B_t$. Then, at a distance $r$ it produces a flux density given by

$$\frac{P_t}{B_t \, 4\pi r^2} . \quad (11)$$

A receiving antenna of effective aperture $A_{er}$ at a distance $r$ can collect a power given by (11) multiplied by both the effective aperture of the receiving antenna and its bandwidth, namely the received power $P_r$ is given by

$$P_r = \frac{P_t}{B_t \, 4\pi r^2} A_{er} B_r . \quad (12)$$

It is assumed that the receiving bandwidth $B_r$ is smaller or, at best (in the "matched bandwidths" case) equal to the transmitting bandwidth $B_t$, that is $B_r \leq B_t$.

So far, we have been talking about an isotropic radiator. But let us now assume that the transmitting antenna has a directivity $D$, that is an antenna gain in the sense of (5):

$$D = \frac{4\pi A_{et}}{\lambda^2} . \quad (13)$$

The received power $P_r$ is then increased by just such a factor due to the directivity of the transmitting antenna, and so (12) must now be replaced by a new equation where the right-hand side is multiplied by such an increased factor, that is

$$P_r = \frac{4\pi A_{et}}{\lambda^2} \cdot \frac{P_t}{B_t \, 4\pi r^2} A_{er} B_r . \quad (14)$$

Rearranging a little, this becomes

$$P_r = \frac{P_t A_{et} A_{er}}{r^2 \lambda^2} \cdot \frac{B_r}{B_t} . \quad (15)$$

This is the *received signal power* expression. For the matched bandwidths case, i.e. for $B_r = B_t$, this is called the Friis transmission formula, since it was first published back in 1946 by the American radio engineer Harald T. Friis (1893-1976) of the Bell Labs. In space missions, we of course know exactly both $B_t$ and $B_r$ and so we may construct our spacecraft so that the two **bands match exactly**, i.e. $B_r = B_t$. Thus, for the case of telecommunications with a spacecraft (but not necessarily for the SETI case) we may well assume the matched bandwidths and have (15) reducing to

$$P_r = \frac{P_t A_{et} A_{er}}{r^2 \lambda^2} . \qquad (16)$$

Let us now rewrite (16) in such a way that we may take into account the gains (i.e. directionalities) of both the transmitting and receiving antennae, that is, in agreement with (5)

$$\begin{cases} G_t = \frac{4\pi A_{et}}{\lambda^2} \\ G_r = \frac{4\pi A_{er}}{\lambda^2} \end{cases} \text{that is} \quad \begin{cases} A_{et} = \frac{G_t \lambda^2}{4\pi} \\ A_{er} = \frac{G_r \lambda^2}{4\pi} \end{cases} \qquad (17)$$

Replacing the last two expressions into (16), we find that (16) is turned into

$$P_r = \frac{P_t G_t G_r}{(4\pi)^2 r^2} \cdot \lambda^2 .$$

This may finally be rewritten in the more traditional form

$$\boxed{P_r = \frac{P_t G_t G_r}{L(r, \lambda)}}. \qquad (18)$$

if one defines

$$\boxed{L(r, \lambda) = (4\pi)^2 \cdot \frac{r^2}{\lambda^2}} \qquad (19)$$

which is the Path Loss (or path attenuation), i.e. the reduction in power density (attenuation) of the electromagnetic waves as they propagates through space. Path loss is a major component in the analysis and design of the link budget of a telecommunication system, see the site
http://www.en.wikipedia.org/wiki/Path_loss .

## 2. BIT ERROR RATE FOR AN "ORDINARY" DIRECT LINK WITH A PROBE AT THE ALPHA CENTAURI DISTANCE

In this section we first define the Bit Error Rate (BER). Then, by virtue of a numerical example, we show that, even at the distance of the nearest star (Alpha Cen. at 4.37 AU) the telecommunications would be impossible by the ordinary powers available today for interplanetary space flight. But in the next section we shall show that the telecommunications would become feasible if we could take advantage of the magnification provided by the Sun's gravity lens, i.e. if we would send out to 550 AU a FOCAL relay spacecraft for each target star system that we wish to communicate with. And this is the key new result presented in this paper.

So, let us start by defining the Bit Error Rate or BER. In telecommunication theory an **error ratio** is the ratio of the number of bits, elements, characters, or blocks incorrectly received to the total number of bits, elements, characters, or blocks sent during a specified time interval. Among these error ratios, the most commonly encountered ratio is the **bit error ratio** (BER) - also called **bit error rate** – that is the number of erroneous bits received divided by the total number of bits transmitted. At the bit error rate Wikipedia site: http://www.en.wikipedia.org/wiki/Bit_error_rate it is shown that the likelihood of a bit misinterpretation

$$p_e = p(0 \mid 1) \, p_1 + p(1 \mid 0) \, p_0. \qquad (20)$$

(believing that we have received a 0 while it was a 1 or the other way round) is basically given by the "complementary error function" or *erfc*(*x*) as follows

$$BER(d, \nu, P_t) = \frac{1}{2} erfc\left(\sqrt{\frac{E_b(d, \nu, P_t)}{N_0}}\right) \qquad (21)$$

In this equation one has:
1) $d$ = distance between the transmitting station on Earth and the receiving antenna in space. For instance, this could be the antenna of a precursor interstellar space probe that was sent out to some light years away.
2) $\nu$ = frequency of the electromagnetic waves used in the telecommunication link. The higher this frequency, the better it is, since the photons are then more energetic ($E = h\nu$). In today's practice, however, the highest $\nu$ for spacecraft links (like the link of the Cassini probe, now at Saturn) are the ones in the Ka band, that is: $\nu_{Ka} \approx 32 \text{ GHz}$ .
3) $P_t$ is the power in watts transmitted by the Earth antenna, typically a NASA Deep Space Network antenna 70 meters in diameter.
4) The complementary error function *ercf*(*x*) is defined by the integral

$$erfc(x) = \frac{2}{\sqrt{\pi}} \int_x^\infty e^{-t^2} dt \qquad (22)$$

(for more maths, see the relevant Wikipedia site: http://www.en.wikipedia.org/wiki/Complementary_error_function ).

5) $E_b(d, \nu, P_t)$ is the received energy per bit, that is the ratio

$$E_b(d, \nu, P_t) = \frac{P_r(d, \nu, P_t)}{\text{Bit\_rate}} \quad (23)$$

6) Finally, $N_0$ is given by the Boltzmann's constant $k$ multiplied by the noise temperature of space far away from the Sun and from any other star. This "empty space noise temperature" might be assumed to equal, say, 100 K.

This is the analytical structure of the MathCad code that this author wrote to yield the BER. Let us now consider the input values that he used in practice:

1) Suppose that a human space probe has reached the Alpha Cen system at 4.37 light year distance from the Sun: then, $d$=4.37 light years.
2) Suppose also that the transmitting antenna from the Earth is a typical NASA Deep Space Network (DSN) antenna having a diameter of 70 meter (like those at Goldstone, Madrid and Canberra), and assume that its efficiency is about 50%.
3) Suppose that the receiving antenna aboard the spacecraft is 12 meter in diameter (it might be an inflatable space antenna, as we supposed in ref. [6] for the FOCAL spacecraft) and assume a 50% efficiency.
4) Suppose that the link frequency is the Ka band (i.e. 32 GHz), as for the Cassini highest frequency.
5) Suppose that the bit rate is 32 kbps = 32000 bit / second. This is the bit rate of ESA's Rosetta interplanetary spacecraft now on its way to a comet.
6) And finally *(this is the most important input assumption) suppose that the transmitting power $P_t$ is moderate: just 40 watts.*

Then:
1) The gain of the transmitting NASA DSN antenna (at this Ka frequency) is about 84 dB.
2) The gain of the spacecraft antenna is about 69 dB.
3) The path loss at the distance of Alpha Cen is 395 dB (a very high indeed path loss with respect to today's interplanetary missions, of course).
4) The power received by the spacecraft at that distance is 2.90 x $10^{-23}$ watt.
5) The received energy per bit (lowered by the noise temperature of the space in between the Sun and Alpha Cen) is 1.3 x $10^{-37}$ joule.

6) *And finally the BER is 0.49, i.e. there is a 50% probability of ERRORS in the telecommunications between the Earth and the probe at Alpha Cen. if we use such a small transmitting power !!!*

In other words, if these are the telecommunication links between the Earth and our probe at Alpha Cen, then this precursor interstellar mission is *worthless*.

The key point in this example is that, for all calculations, (18) and (19) were used *WITHOUT TAKING THE GAIN OF THE SUN GRAVITY LENS INTO ACCOUNT, because this was a DIRECT link and NOT a FOCAL mission.*

## 3. BIT ERROR RATE AT THE ALPHA CENTAURI DISTANCE ENHANCED BY THE MAGNIFICATION PROVIDED BY THE SUN'S GRAVITY LENS (FOCAL)

The disappointing BER results of the previous section are totally reversed, however, if we suppose that a FOCAl space mission has been previously sent out to 550 AU in the direction opposite to Alpha Cen so that we now have the MAGNIFICATION of the Sun's Gravity Lens playing in the game.

*Mathematically, this means that we must introduce a third multiplicative gain at the numerator of (**18**): the Sun's Gravity Lens GAIN, given by (**8**) where the Schwarzschild radius of the Sun is given by (**2**).*

This new gain is huge at the Ka band frequency:

$$G_{Sun}(\nu_{Ka}) = 12444837 \sim 70 \text{ dBs} \quad (24)$$

and so the received power (18) at Alpha Cen, with the usual Earth-transmitted power of just 40 watts becomes

$$P_r = 2.9 \times 10^{-23} \text{ watts} \quad (25)$$

and the relevant BER becomes absolutely acceptable:

$$\boxed{\text{BER} = 0.000000526387845} \quad (26)$$

*This should convince anybody that the FOCAL space mission is indispensable to keep the link at interstellar distances equal or higher than Alpha Cen.*

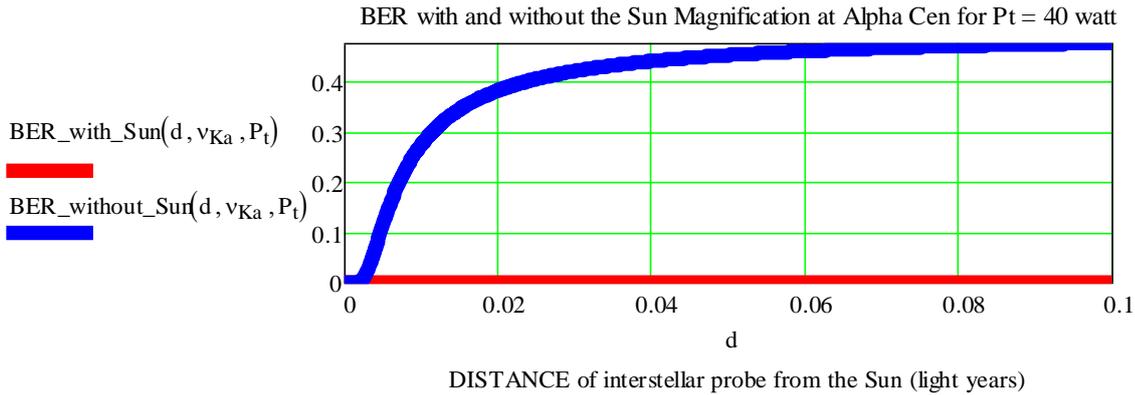

Figure 3. The Bit Error Rate (BER) (upper, blue curve) tends immediately to the 50% value (BER = 0.5) even at moderate distances from the Sun (0 to 0.1 light years) for a 40 watt transmission from a DSN antenna that is a DIRECT transmission, i.e. without using the Sun's Magnifying Lens. On the contrary (lower red curve) the BER keeps staying at zero value (perfect communications!) if the FOCAL space mission is made, so as the Sun's magnifying action is made to work.

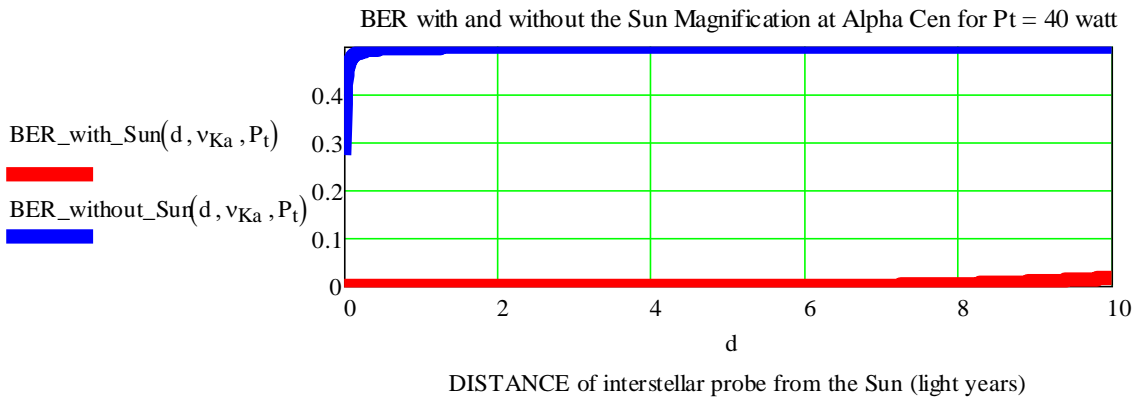

Figure 4. Same as in Figure 3, but for probe distances up to 10 light years. We see that at about 9 light years away the BER curve starts being no exactly flat any more, and starts increasing slowly.

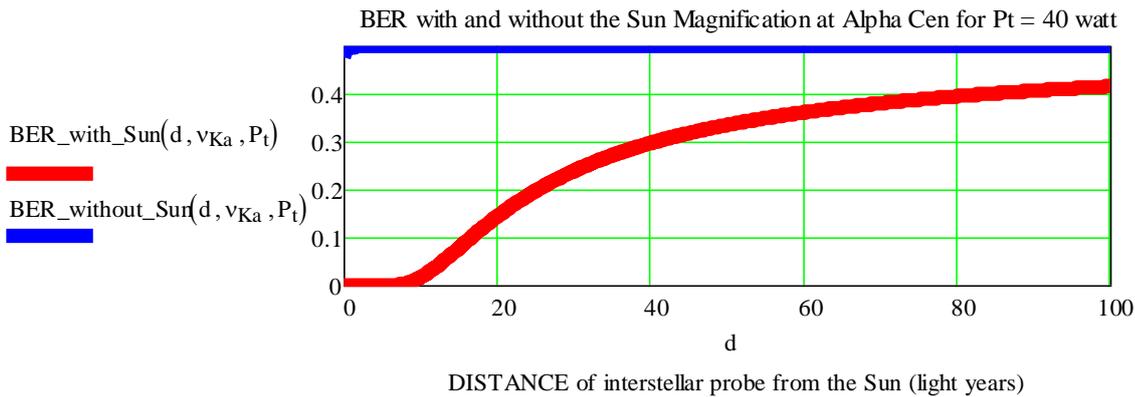

Figure 5. Same as in Figure 4, but for probe distances up to 100 light years. We see that, from 9 light years onward, the Sun-BER increases, reaching the dangerous level of 40% (Sun-BER = 0.4) at about 100 light years. Namely, at 100 light years even the Sun's Lens cannot cope with this very low transmitted power of 40 watt.

## 4. IN THIS SECTION WE FIRSTLY DEFINE THE INTERSTELLAR RADIO BRIDGES FOR CHEAP INTERSTELLAR COMMUNICATIONS

In three published papers (refs. [3], [4] and [5]) this author mathematically described the "radio bridges" created by the gravitational lens made up by the Sun plus any nearby star like Alpha Centauri A, or Barnard's star, or Sirius. The result is that it is indeed possible to communicate between the solar system and a nearby stellar system with modest signal powers if two FOCAL missions are set up:
1) One by Humans at least at 550 AU from the Sun in the opposite direction to the selected star, and
2) One by ETs at the minimal focal distance of their own star in the direction opposite to the Sun.

Far-sighted readers should realize that a Civilization much more advanced than Humans in the Galaxy might already have created such a network of cheap interstellar radio links: a true *Galactic Internet* that Humans will be unable to access as long as they won't have access to the magnifying power of their own star, the Sun, i.e. until they will be able to reach the minimal focal distance of 550 AU by virtue of their own FOCAL space missions. Figure 6 hereafter shows the interstellar radio bridge concept, with the two stars acting as two magnifying lenses and the two FOCAL spacecrafts located on opposite sides and beyond the minimal focal distance of each star.

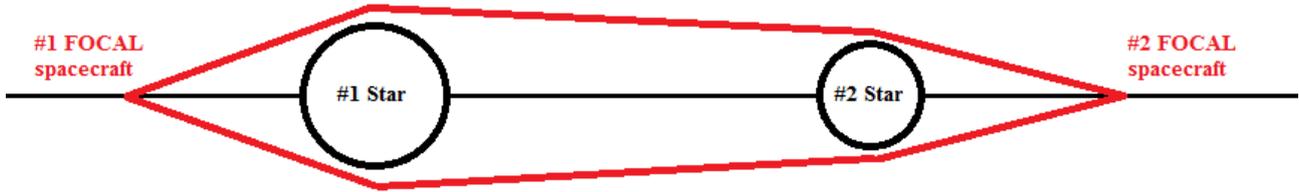

**Figure 6. RADIO BRIDGE between any two stars.** The electromagnetic waves path (red, solid lines) are deflecteded by the gravity field of each star, and made to focus on the two FOCAL spacecrafts placed on opposite sides with respect to the two stars. These two FOCAL spacecrafts must be strictly positioned along the axis passing thorugh the two stars' centers. Then, the trasmission powers between the #1 FOCAL spacecraft and the #2 FOCAL spacecraft are greatly reduced. In fact, the huge (antenna) gains of the two stars, plus the modest (antenna) gains of the two FOCAL spacecrafts combine to yield a huge total (antenna) gain of the whole system, meaning that the trasmissions powers between the two stellar systems become quite affordable. This is the key to build a GALACTIC INTERNET, that might already have been created in the Galaxy by Aliens capable of sending FOCAL spacecrafts around. But we, Humans, still are cut off from all that since we did not yet send any FOCAL spacecraft 550 AU away from the Sun, namely we have not yet learned about how to use the Sun as a gravitational magnifying lens.

## 5. THE FANTASTIC "RADIO BRIDGE" BETWEEN THE SUN AND ALPHA CEN A BY USING THE TWO GRAVITATIONAL LENSES OF BOTH JUST MATCHED TO EACH OTHER

In this section we provide one more new result: we define the radio bridge between the Sun and Alpha Cen A by using BOTH gravitational lenses!
In other words, suppose that in future we will be able to send a probe to Alpha Cen A and suppose that we succeed in placing this probe just on the other side of Alpha Cen A with respect to the Sun and at the minimal focal distance typical of Alpha Cen A. This distance is NOT 550 AU obviously because both the radius and the mass of Alpha Cen A are different (actually slightly higher) than the values of the Sun:

$$\begin{cases} r_{Alpha\_Cen\_A} = 1.227 \, r_{Sun} \\ M_{Alpha\_Cen\_A} = 1.100 \, M_{Sun} \end{cases} \quad (27)$$

Replacing these values into (1) (obviously rewritten for Alpha Cen A), the relevant minimal focal distance is found

$$d_{focal\_Alpha\_Cen\_A} \approx \frac{c^2 r^2_{Alpha\_Cen\_A}}{4GM_{Alpha\_Cen\_A}} \approx 749 \text{ AU}. \quad (28)$$

The Schwarzschild radius for Alpha Cen A is given by

$$r_{Schwarzschild\_Alpha\_Cen\_A} = \frac{2GM_{Alpha\_Cen\_A}}{c^2} = 3.248 \text{ km}$$
...(29)

And so the gain, provided by (8), turns out to equal

$$G_{Alpha\_Cen\_A}(\nu_{Ka}) = 4\pi^2 \frac{r_{Schwarzschild\_Alpha\_Cen\_A}}{\lambda_{Ka}} =$$

$$= 13689321 .\qquad(30)$$

That is

$$\boxed{G_{Alpha\_Cen\_A}(\nu_{Ka}) \approx 71 \text{ dB}}\qquad(31)$$

Incidentally, we chose Alpha Cen A, and not B or C, because it has the highest mass, and so the highest gain, in the whole Alpha Cen triple system. The future telecommunications between the Sun and the Alpha Cen system are thus optimized by selecting Alpha Cen A as the star on the other side of which to place a FOCAL spacecraft at the minimal distance of 750 AU. That FOCAL spacecraft would then easily relay its data anywhere within the Alpha Cen system.

Having found the Alpha Cen A gain (31) we are now able to write the new equation corresponding to (18) for the Sun-Alpha Cen bridge. In fact, we must now put at the numerator of (18) three gains:
1) The Sun gain at 32 GHz,
2) The Alpha Cen A gain at 32 GHz, and
3) The 12-meter FOCAL antenna gain at 32 GHz raised to the square because there are two such 12-meter antennas: one at 550 AU from the Sun and one at 749 AU from Alpha Cen A, and they must be perfectly aligned with the axis passing thru both the Sun and Alpha Cen A.

Thus, the received power given by (18) now reads

$$P_r = \frac{P_t \, G_{Sun} \, G_{Alpha\_Cen\_A} \left(G_{12\_meter\_antenna\_at\_Ka}\right)^2}{L(r,\lambda)} .$$
...(32)

where obviously $r$ equals 4.37 light years and $\lambda$ corresponds to a 32 GHz frequency.

Let us now go back to the BER and replace (18) with (32) in the long chain of calculations described in Section 4. Since the received power $P_r$ has now changed, clearly both (23) and (21) yield different numerical results. But now:
1) The link frequency has been fixed at 32 GHz (Ka band), and so no longer is an independent variable in the game.
2) Also the distance $d$ has been fixed (it is the distance of Alpha Cen A) and so is no longer an independent variable in the game.
3) It follows that, in (23) and (21), the only variable to be free to vary is now the transmitted power, $P_t$.

Let us re-phrase the last sentence in different terms. Practically, we are now studying the BER as a function of the transmitted power $P_t$ only and, physically, this means that:
a) We start by inputting very low transmission powers in watts, and find out that the BER is an awful 50%, i.e. the telecommunication between the Sun and Alpha Cen is totally disrupted. This is of course because the energy per bit is so much lower than the empty space noise temperature.
b) We then increase the transmitted power, at a certain point the BER starts getting smaller than 50%. And so it gets smaller and smaller until the transmitted power is so high that the BER gets down to zero and the telecommunications are just perfect.
c) But the surprise is that… for the Sun-Alpha Cen direct radio bridge exploiting both the two gravitational lenses, this minimum transmitted power is incredibly… small ! Actually it just equals less than $10^{-4}$ watts, i.e. one tenth of a milliwatt is enough to have perfect communication between the Sun and Alpha Cen through two 12-meter FOCAL spacecraft antennas. How is that possible?
d) Well, that is the "miracle' given to Humanity by the Gravitational Lenses to both explore the universe and keep the link with other stars, you know! Just remember that, in 2009, the discovery of the first extrasolar planet in the Andromeda galaxy (M31) was announced because of the gravitational lens caused by something in between!

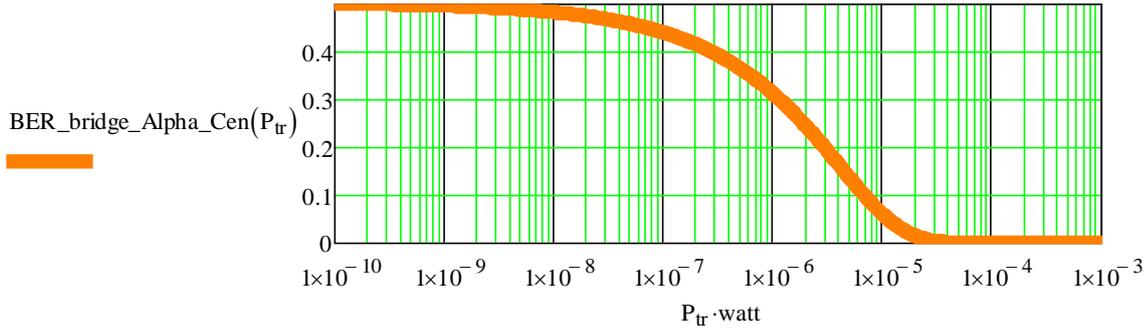

**Figure 6. Bit Error Rate (BER) for the double-gravitational-lens system giving the radio bridge between the Sun and Alpha Cen A.** In other words, there are two gravitational lenses in the game here: the Sun one and the Alpha Cen A one, and two 12-meter FOCAL spacecrafts are supposed to have been put along the two-star axis on opposite sides at or beyond the minimal focal distances of 550 AU and 749 AU, respectively. This radio bridge has an OVERALL GAIN SO HIGH that a miserable $10^{-4}$ watt transmitting power is sufficient to let the BER get down to zero, i.e. to have perfect telecommunications! Fantastico!
Notice also that the scale of the horizontal axis is logarithmic, and the trace is yellowish since the light of Alpha Cen A is yellowish too. This will help us to distinguish this curve from the similar curve for the Barnard's Star, that is a small red star 6 light years away, as we study in the next section.

### 6. THE "RADIO BRIDGE" BETWEEN THE SUN AND BARNARD'S STAR USING THE TWO GRAVITATIONAL LENSES OF BOTH

The next closest star to the Sun beyond the triple Alpha Cen system is Barnard's star (see, for instance, t http://www.en.wikipedia.org/wiki/Barnard's_Star ). Let us now repeat for the gravitational lens of Barnard's star the same calculations that we did in the previous section for Alpha Cen A. Then one has:

$$\begin{cases} d_{Barnard} = 5.98 \text{ light years} \\ r_{Barnard} = 0.17\, r_{Sun} \\ M_{Barnard} = 0.16\, M_{Sun}. \end{cases} \quad (33)$$

Barnard's star is thus just a small red star, that is actually "passing by" the Sun right now and is not known to have planets around it, As a consequence of the numbers listed in (33), one infers that

$$\begin{cases} d_{focal\_Barnard} = 98 \text{ AU} \\ r_{Schwarschild\_Barnard} = 0.47 \text{ km} \\ G_{Barnard}(\nu_{Ka}) = 1991174. \end{cases} \quad (34)$$

Especially the gain is important to us:

$$G_{Barnard}(\nu_{Ka}) = 63 \text{ dB}. \quad (35)$$

We replace this into the Barnard's star equivalent of (32), again supposing that two 12-meter FOCAL spacecraft antennas are placed along the Sun-Barnard straight line at or beyond 550 AU and 100 AU, respectively. The result is the new graph of the BER as a function of the transmitted power only as in Figure 7.

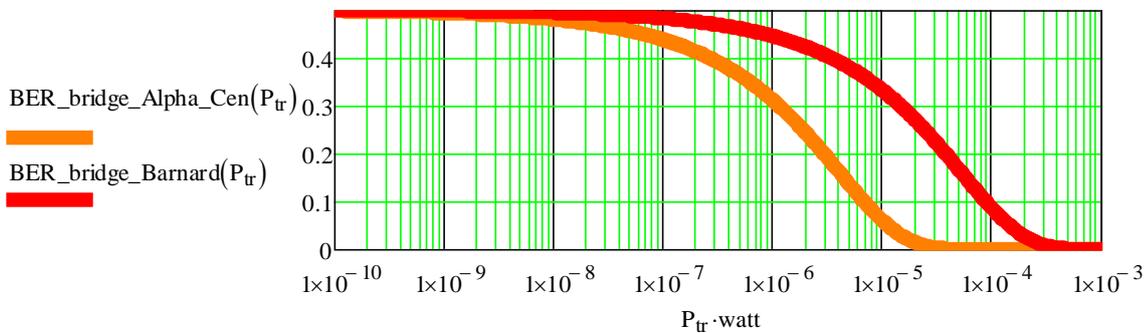

**Figure 7. Bit Error Rate (BER) for the double-gravitational-lens of the radio bridge between the Sun and Alpha Cen A (yellowish curve) plus the same curve for the radio bridge between the Sun and Barnard's star (reddish curve, just as Barnard's star is a reddish star): for it, $10^{-3}$ watt are needed to keep the BER down to zero, because the gain of Barnard's star is so small when compared to the Alpha Centauri A's.**

## 7. THE "RADIO BRIDGE" BETWEEN THE SUN AND SIRIUS-A USING THE TWO GRAVITATIONAL LENSES OF BOTH

The next star we want to consider is Sirius A. This is because Sirius A is a big, massive bluish star and so it is completely different from both Alpha Cen A (which is a Sun-like star) and Barnard's star, which is a small red star. Data may again be taken from the Wikipedia site:
http://www.en.wikipedia.org/wiki/Sirius and one gets:

$$\begin{cases} d_{Sirius\_A} = 8.6 \text{ light years} \\ r_{Sirius\_A} = 1.711 \, r_{Sun} \\ M_{Sirius\_A} = 2.02 \, M_{Sun}. \end{cases} \quad (36)$$

From these data one gets:

$$\begin{cases} d_{focal\_Sirius\_A} = 793 \text{ AU} \\ r_{Schwarschold\_Sirius\_A} = 5.96 \text{ km} \\ G_{Sirius\_A}(\nu_{Ka}) = 251385723 \end{cases} \quad (37)$$

The important thing is of course the gain:

$$G_{Sirius\_A}(\nu_{Ka}) = 74 \text{ dB.} \quad (38)$$

Then, one places this into the Sirius A equivalent of (32), again supposing that two 12-meter FOCAL spacecraft antennas are placed along the Sun-Sirius A straight line at or beyond 550 AU and 793 AU, respectively. The result is the new graph of the BER as a function of the transmitted power only as in Figure 8.

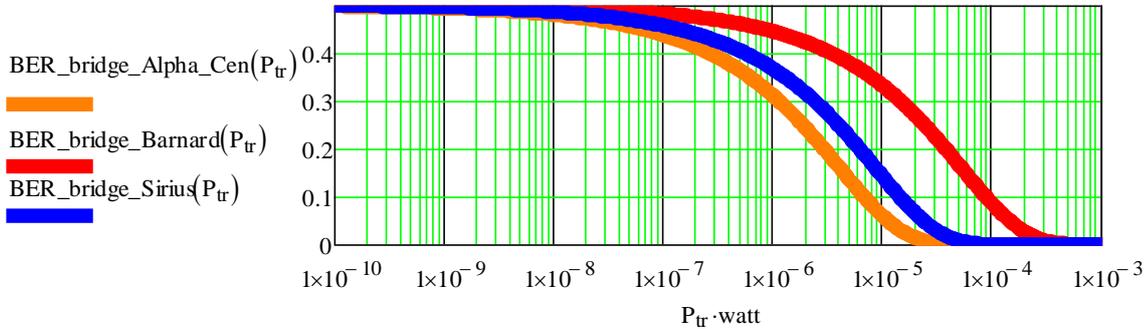

**Figure 8.** Bit Error Rate (BER) for the double-gravitational-lens of the radio bridge between the Sun and Alpha Cen A (yellowish curve) plus the same curve for the radio bridge between the Sun and Barnard's star (reddish curve, just as Barnard's star is a reddish star) plus the same curve of the radio bridge between the Sun and Sirius A (blue curve, just as Sirius A is a big blue star). From this blue curve we see that only $10^{-4}$ watt are needed to keep the BER down to zero, because the gain of Sirius A is so big when compared the gain of the Barnard's star that it "jumps closer to Alpha Cen A's gain" even if Sirius A is so much further out than the Barnard's star! In other words, the star's gain and its size combined matter even more than its distance!

## 8. THE "RADIO BRIDGE" BETWEEN THE SUN AND ANOTHER SUN-LIKE STAR LOCATED AT THE GALACTIC BULGE USING THE TWO GRAVITATIONAL LENSES OF BOTH

Tempted by the suggestion to increase the distance of the second star more and more, and then see what our calculations yield, we now imagine that the second star is Sun-like (i.e. that it has the same radius and mass exactly as the Sun) but is located… inside the Galactic Bulge! Namely 26,000 light years away, according to the Wikipedia "Milky Way Galaxy" site:
http://www.en.wikipedia.org/wiki/Milky_Way . So, the equivalent of (18) and (32) now becomes

$$P_r = \frac{P_t \, (G_{Sun\_at\_Ka})^2 (G_{12\_meter\_antenna\_at\_Ka})^2}{L(r,\lambda)} . \quad …(39)$$

and the plot of the BER vs. transmitted power is shown in Figure 9 as the new, pink curve at the far right of the previous three curves of Alpha Cen A (orangish), Barnard's star (red) and Sirius A (blu). The new pink curve showing the BER of a Sun-like star at the Galactic Bulge is naturally much to the right of the previous three stellar curves inasmuch as the Bulge distance of 26,000 light is so much higher than the distances of the three mentioned nearby stars (all less than 10 light years away anyway). The horizontal axis scale is much higher now, since the pink BER curve gets to zero only for transmitted power of about 1000 watt.

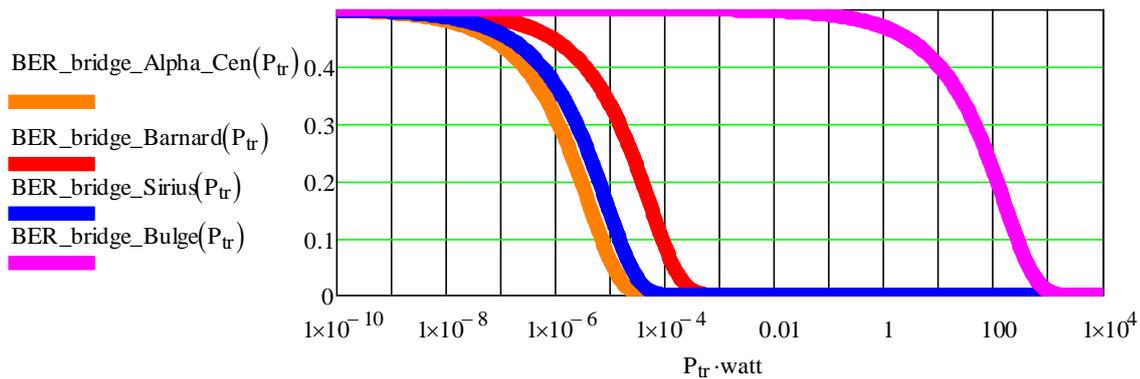

**Figure 9. Bit Error Rate (BER) for the double-gravitational-lens of the radio bridge between the Sun and Alpha Cen A (orangish curve) plus the same curve for the radio bridge between the Sun and Barnard's star (reddish curve, just as Barnard's star is a reddish star) plus the same curve of the radio bridge between the Sun and Sirius A (blue curve, just as Sirius A is a big blue star). In addition, to the far right we now have the pink curve showing the BER for a radio bridge between the Sun and another Sun (identical in mass and size) located inside the Galactic Bulge at a distance of 26,000 light years. The radio bridge between these two Suns works and their two gravitational lenses works perfectly (i.e. BER = 0) if the transmitted power is higher than about 1000 watts.**

## 9. THE "RADIO BRIDGE" BETWEEN THE SUN AND ANOTHER SUN-LIKE STAR LOCATED INSIDE THE ANDROMEDA GALAXY (M 31) USING THE TWO GRAVITATIONAL LENSES OF BOTH

We conclude this paper by calculating the radio bridge between the Sun and another Sun… in Andromeda! The distance is now 2.5 million light year, but the bridge would still work if the transmitted power was of higher than about $10^7$ watts = 10 Megawatt. This is shown by the new cyan curve on the far right in Figure 10 below. Perhaps this idea is not as "crazy" as it might appear, if recently (June 2009) the first extrasolar planet in the Andromeda galaxy was announced to have been discovered just by gravitational lensing. See, for instance, the web site:
http://www.redorbit.com/.../possible_planet_found_outside_our_galaxy/index.html

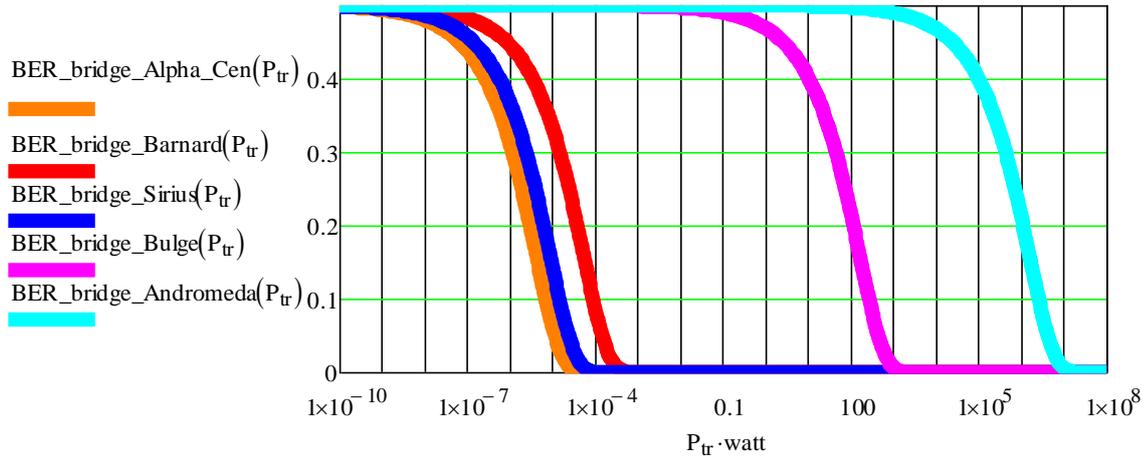

**Figure 10.** The same four Bit Error Rate (BER) curves as shown in Figure 9 plus the new cyan curve appearing here on the far right: this is the BER curve of the radio bridge between the Sun and another Sun just the same but located somewhere in the Andromeda Galaxy M 31. Notice that this radio bridge would work fine (i.e. with BER = 0) if the transmitting power was at least $10^7$ watt = 10 Megawatt. This is not as "crazy" at it might seem if one remembers that recently (June 2009) the discovery of the first extrasolar planet in the Andromeda Galaxy was announced, and the method used for the detection was just GRAVITATIONAL LENSING !

## 10. BIT ERROR RATE AND CHANNEL CAPACITY

We now want to increas the number of stars that we will calculate the Bit Errore Rate (BER) for, together with the channel capacity, if we envisage a communication channel between those stars and our Sun.

Let's make use of the following quantities:
- The Boltzmann constant, $k$, $1.38064852 \cdot 10^{-23}$ m$^2$kgs$^{-2}$K$^{-1}$;
- The gravitational constant, $G$, $6.67408 \cdot 10^{-11}$ m$^3$kg$^{-1}$s$^{-2}$;
- The speed of light, $c$, 299792458 ms$^{-1}$;
- The mass of the Sun, $M_{Sun}$, $1,989 \cdot 10^{30}$ kg;
- The radius of the Sun, $r_{Sun}$, $6.96340 \cdot 10^{30}$ m;
- The frequency of the transmission, $\nu$, $32 \cdot 10^9$ Hz;
- The bit rate, $32 \cdot 10^4$ bits$^{-1}$;
- The efficiency of the spacecraft antennas at the Sun and star foci, $p_{eff}$, 50%;
- The radius of the transmitting antenna, $r_t$, 6 m;
- The radius of the receiving antenna, $r_r$, 6 m;
- The noise temperature in deep space, $T$, 100 K;
- The transmission power, $P_t$, $4 \cdot 10^{-6}$ W;
- The communication bandwidth, $W$, $5 \cdot 10^8$ Hz.

We have selected the 50 stars nearest to our Sun (https://it.wikipedia.org/wiki/Stelle_pi%C3%B9_vicine_alla_Terra). We have searched for their:
- Distance from the Sun, $d_{star}$ (m);
- Mass of the star, $M_{star}$ (kg);
- Radius of the star, $r_{star}$ (m).

Then we calculated:
- The focal distance of the star, $d_{fstar}$ (m), using the equation (1);
- The bridge gain given by the gains of the receiving and transmittin gantennas, together with the lens antenna of the Sun and the star, namely $G_{star}G_{ta}G_{ra}G_{Sun}$, see the equations (7) and (10);
- The received power, $P_r$, using the equation (18);
- The energy per bit, $E_b$, using the equation (23);
- The Bit Error Rate (BER), using the equation (21);
- The channel capacity, $C$, given by th equation

$$C(W) = W \log_2\left(1 + \frac{P_r}{WkT}\right)$$

In the table 1, the stars are sorted by the ascending BER and descending channel capacity.

For the first 6 of the stars in Table 1, we plot the channel capacity as a function of the transmitted power, see the Figure 11. We suppose the communication is between antennas at the foci of the Sun and the stars.

Thanks to the antenna gain due to the lensing effect of the Sun + star system, the channel capacity is very high; all the channel have capacity very close to each other (the scale is logarithmic), therefore, we provide a detail in Figure 12.

## CONCLUSION

As of September 29th, 2021, the FOCAL space mission turned out to be of interest to NIAC, the NASA Institute for Advanced Concepts - Wikipedia that in 2020 funded its further study by virtue of a $2million grant assigned to a JPL team led by Dr. Slava Turyshev. However, here the envisaged FOCAL mission goal is the observation of exoplanets, rather than keeping the radio link across interstellar distance. And something similar might be true for China also, though we have no info at all about that.

Thus, a much longer time (centuries or millenia) might be necessary for the creation of interstellar radio bridges to just the closest stars to the Sun: the three Alpha Cen stars, and "big" Sirius, much more than just "small" Barnard's Star.

Yet we believe that the present paper will still be useful in a few centuries to set up the ROADMAP of the Human Expansion into the Galaxy. In fact, it makes it clear that keeping the radio link among the Solar system and "nearby" stellar systems is crucial to the Human effort to "colonize" those systems, and the present paper just aims at that.

Thanks to those who will make use of our research work in the future and give us the credit for the new results provided by this paper.

## ACKNOWLEDGMENTS

The two authors are indebted to many colleagues for conversations and suggestions, but, in particular, would like to thank Paul Gilster for maintaining his terrific Centauri Dreams web site: http://www.centauri-dreams.org .

## REFERENCES

[1] C. Maccone, "*Realistic targets at 1000 AU for interstellar precursor missions*", *Acta Astronautica*, Vol. 67 (2010), pages 526-538.

[2] C. Maccone, "*Interstellar Radio Links Enhanced by Exploiting the Sun as a Gravitational Lens*", paper #IAC-09.D4.1.8 presented by the author at the 60th International Astronautical Congress held at Daejeon, Republic of Korea, October 12th-16th, 2009, and distributed to all participants as a CD-ROM file, but not published in a printed form.

[3] C. Maccone, "*Interstellar radio links enhanced by exploiting the Sun as a Gravitational Lens*", *Acta Astronautica*, Vol. 68, (2011), pages 76-84.

[4] C. Maccone, "*Galactic Internet made possible by star gravitational lensing*", *Acta Astronautica*, Vol. 82, (2013), pages 246-250.

[5] C. Maccone, "*Going to Alpha Centauri B and setting up a Radio Bridge*", *Acta Astronautica*, Vol. 104, (2014), pages 458-463.

[6] C. Maccone, "*Deep Space Flight and Communications – Exploiting the Sun as a Gravitational Lens*", a 400-pages treatise about the FOCAL space mission that embodies and updates all previously published material about FOCAL. ISBN 978-3-540-72942-6, published by Springer, Berlin, Heidelberg, New York, 2009. Library of Congress Control Number: 2007939976. © Praxis Publishing Ltd, Chichester, UK, 2009.

[7] John D. Kraus, *Radio Astronomy*, 2nd ed, Cygnus-Quasar Books, Powell, Ohio, 1966, pp. **6**-115 thru **6**-118.

[8] R. Orta, P. Savi and R. Tascone, "Analysis of Gravitational Lens Antennas", in *Proceedings of the International Conference on Space Missions and Astrodynamics* held in Turin, Italy, June 18, 1992, C. Maccone editor, *Journal of the British Interplanetary Society*, Vol. 47 (1994), pages 53-56.

| Star | Distance from the Sun ($d_{star}$) (m) | Mass of the star ($M_{star}$) (kg) | Radius of the star ($r_{star}$) (m) | Focal distance of the star ($d_{fstar}$) (m) | Bridge gain | Received power ($P_r$) (W) | Energy per bit ($E_b$) (Wbit$^{-1}$s) | Bit Error Rate (BER) | Channel capacity (C) bits$^{-1}$ | Channel capacity (C) for infinite bandwidth bits$^{-1}$ |
|---|---|---|---|---|---|---|---|---|---|---|
| Proxima Centauri | 4,00158E+16 | 2,193E+29 | 100775000 | 1,55904E+13 | 3,35748E+35 | 4,66157E-10 | 1,45674E-14 | 0 | 4700730818 | 4,87106E+11 |
| Barnard star | 5,61924E+16 | 3,14E+29 | 136000000 | 1,98307E+13 | 4,80734E+35 | 3,38479E-10 | 1,05775E-14 | 0 | 4470257166 | 3,5369E+11 |
| Wolf 359 | 7,35988E+16 | 1,7901E+29 | 132240000 | 3,28881E+13 | 2,74064E+35 | 1,12484E-10 | 3,51514E-15 | 0 | 3678531496 | 1,1754E+11 |
| Lalande 21185 | 7,84234E+16 | 8,954E+29 | 315000000 | 3,73073E+13 | 1,37086E+36 | 4,95544E-10 | 1,54858E-14 | 0 | 4744766374 | 5,17814E+11 |
| Sirius A | 8,11668E+16 | 4,1769E+30 | 1309119200 | 1,38132E+14 | 6,39483E+36 | 2,15801E-09 | 6,74379E-14 | 0 | 5805301550 | 2,25499E+12 |
| BL Ceti | 8,2302E+16 | 1,989E+29 | 97487600 | 1,60862E+13 | 3,04516E+35 | 9,99472E-11 | 3,12335E-15 | 0 | 3593839510 | 1,04439E+11 |
| Ross 154 | 9,16674E+16 | 3,1824E+29 | 125341200 | 1,66197E+13 | 4,87225E+35 | 1,28909E-10 | 4,02839E-15 | 0 | 3776281675 | 1,34702E+11 |
| Ross 248 | 9,76272E+16 | 3,1824E+29 | 83560800 | 7,38653E+12 | 4,87225E+35 | 1,1365E-10 | 3,55157E-15 | 0 | 3685922956 | 1,18758E+11 |
| Epsilon Eridani | 9,933E+16 | 1,63098E+30 | 515291600 | 5,48084E+13 | 2,49703E+36 | 5,62658E-10 | 1,75831E-14 | 0 | 4836268782 | 5,87944E+11 |
| Lacaille 9352 | 1,01222E+17 | 9,7461E+29 | 327279800 | 3,69996E+13 | 1,49213E+36 | 3,23771E-10 | 1,01178E-14 | 0 | 4438278130 | 3,38321E+11 |
| Ross 128 | 1,04911E+17 | 3,34E+29 | 136400000 | 1,87531E+13 | 5,11354E+35 | 1,0329E-10 | 3,22781E-15 | 0 | 3617409716 | 1,07932E+11 |
| EZ Aquarii A | 1,0652E+17 | 1,989E+29 | 146231400 | 3,61940E+13 | 3,04516E+35 | 5,96668E-11 | 1,86459E-15 | 0 | 3225052686 | 62348202939 |
| Procyon A | 1,07844E+17 | 2,93974E+30 | 1426104320 | 2,32907E+14 | 4,50074E+36 | 8,60348E-10 | 2,68859E-14 | 0 | 5142293597 | 8,99012E+11 |
| 61 Cygni A | 1,07466E+17 | 1,3923E+30 | 463066100 | 5,18493E+13 | 2,13161E+36 | 4,10346E-10 | 1,28233E-14 | 0 | 4608888804 | 4,28787E+11 |
| Gliese 725 A | 1,07749E+17 | 6,5637E+29 | 243719000 | 3,04663E+13 | 1,0049E+36 | 1,92431E-10 | 6,01348E-15 | 0 | 4064010059 | 2,01079E+11 |
| GX Andromedae | 1,09925E+17 | 7,5582E+29 | 236755600 | 2,49673E+13 | 1,15716E+36 | 2,12902E-10 | 6,6532E-15 | 0 | 4136686463 | 2,2247E+11 |
| Epsilon Indi A | 1,11723E+17 | 1,49175E+30 | 508328200 | 5,83153E+13 | 2,28387E+36 | 4,0679E-10 | 1,27122E-14 | 0 | 4602621101 | 4,25072E+11 |
| DX Cancri | 1,11817E+17 | 1,73043E+29 | 76597400 | 1,14147E+13 | 2,64929E+35 | 4,71079E-11 | 1,47212E-15 | 0 | 3056769299 | 49224920215 |
| Tau Ceti | 1,12574E+17 | 1,55142E+30 | 550108600 | 6,56686E+13 | 2,37522E+36 | 4,16687E-10 | 1,30215E-14 | 0 | 4619931231 | 4,35413E+11 |
| Gliese 1061 | 1,13236E+17 | 2,3868E+29 | 108629040 | 1,66443E+13 | 3,65419E+35 | 6,33581E-11 | 1,97994E-15 | 0 | 3267872729 | 66205438534 |
| YZ Ceti | 1,15507E+17 | 2,82438E+29 | 109325380 | 1,42465E+13 | 4,32412E+35 | 7,20554E-11 | 2,25173E-15 | 0 | 3359721895 | 75293543720 |
| Luyten star | 1,17304E+17 | 5,7681E+29 | 201938600 | 2,38010E+13 | 8,83096E+35 | 1,4268E-10 | 4,45876E-15 | 0 | 3849129658 | 1,49092E+11 |
| Teegarden star | 1,17872E+17 | 1,77021E+29 | 74508380 | 1,05578E+13 | 2,71019E+35 | 4,33674E-11 | 1,35523E-15 | 0 | 2997988246 | 45316320884 |
| Kapteyn star | 1,2071E+17 | 5,5692E+29 | 201938600 | 2,46511E+13 | 8,52644E+35 | 1,30097E-10 | 4,06552E-15 | 0 | 3782864288 | 1,35943E+11 |
| Lacaille 8760 | 1,2175E+17 | 1,19346E+30 | 459584400 | 5,95817E+13 | 1,82719E+36 | 2,74047E-10 | 8,56397E-15 | 0 | 4318281928 | 2,86363E+11 |
| Kruger 60 A | 1,24399E+17 | 5,39019E+29 | 243719000 | 3,70992E+13 | 8,25238E+35 | 1,18557E-10 | 3,70491E-15 | 0 | 3716233328 | 1,23885E+11 |
| Ross 614 A | 1,25818E+17 | 4,3758E+29 | 174085000 | 2,33161E+13 | 6,69935E+35 | 9,40868E-11 | 2,94021E-15 | 0 | 3550560470 | 98315076222 |
| Gliese 628 | 1,3244E+17 | 5,7681E+29 | 215865400 | 2,71971E+13 | 8,83096E+35 | 1,11931E-10 | 3,49785E-15 | 0 | 3674996474 | 1,16961E+11 |
| Gliese 35 | 1,33008E+17 | 1,3923E+30 | 9052420 | 1,98147E+10 | 2,13161E+36 | 2,67878E-10 | 8,37118E-15 | 0 | 4301899074 | 2,79916E+11 |
| Gliese 1061 | 1,33859E+17 | 8,9505E+29 | 334243200 | 4,20211E+13 | 1,37032E+36 | 1,70023E-10 | 5,31323E-15 | 0 | 3975044989 | 1,77664E+11 |
| Wolf 424 A | 1,35278E+17 | 2,5857E+29 | 111414400 | 1,61620E+13 | 3,95871E+35 | 4,80929E-11 | 1,5029E-15 | 0 | 3071482946 | 50254157756 |
| TZ Arietis | 1,38116E+17 | 1,77021E+29 | 348170000 | 2,30541E+14 | 2,71019E+35 | 3,15859E-11 | 9,8706E-16 | 0 | 2773527486 | 33005401217 |
| Gliese 867 | 1,39724E+17 | 8,1549E+29 | 292462800 | 3,53112E+13 | 1,24851E+36 | 1,42178E-10 | 4,44306E-15 | 0 | 3846597729 | 1,48567E+11 |
| LHS 292 | 1,40008E+17 | 1,5912E+29 | N/A | N/A | 2,43613E+35 | 2,76271E-11 | 8,63428E-16 | 0 | 2679201768 | 28871362363 |
| Gliese 674 | 1,40103E+17 | N/A | N/A | N/A | N/A | N/A | N/A | N/A | N/A | N/A |
| Gliese 1245 A | 1,40103E+17 | N/A | N/A | N/A | N/A | N/A | N/A | N/A | N/A | N/A |
| Gliese 440 | 1,42468E+17 | N/A | N/A | N/A | N/A | N/A | N/A | N/A | N/A | N/A |
| Gliese 1002 | 1,44833E+17 | N/A | N/A | N/A | N/A | N/A | N/A | N/A | N/A | N/A |
| Ross 780 | 1,45116E+17 | 6,9615E+29 | 250682400 | 3,03903E+13 | 1,06581E+36 | 1,12519E-10 | 3,51623E-15 | 0 | 3678752724 | 1,17576E+11 |
| Gliese 412 A | 1,49468E+17 | 9,5472E+29 | 264609200 | 2,46902E+13 | 1,46168E+36 | 1,45458E-10 | 4,54555E-15 | 0 | 3862969653 | 1,51994E+11 |
| Groombridge 1618 | 1,50225E+17 | 1,27296E+30 | 452621000 | 5,41807E+13 | 1,9489E+36 | 1,91994E-10 | 5,99982E-15 | 0 | 4062375704 | 2,00623E+11 |
| Gliese 388 | 1,50792E+17 | N/A | N/A | N/A | N/A | N/A | N/A | N/A | N/A | N/A |
| LHS 288 | 1,50792E+17 | N/A | N/A | N/A | N/A | N/A | N/A | N/A | N/A | N/A |
| Gliese 832 | 1,52306E+17 | 8,9505E+29 | 334243200 | 4,20211E+13 | 1,37032E+36 | 1,31332E-10 | 4,10412E-15 | 0 | 3789645204 | 1,37234E+11 |
| LP 944-020 | 1,53157E+17 | 1,3923E+29 | 174085000 | 7,32791E+13 | 2,13161E+35 | 2,02029E-11 | 6,31341E-16 | 0 | 2459808483 | 21110818109 |
| DENIS/DEN 02554700 | 1,53252E+17 | N/A | N/A | N/A | N/A | N/A | N/A | N/A | N/A | N/A |
| Gliese 682 | 1,5136E+17 | 5,3703E+29 | N/A | N/A | 8,22193E+35 | 7,97872E-11 | 2,49335E-15 | 0 | 3432583780 | 83372825549 |
| 40 Eridani A | 1,5382E+17 | 1,53153E+30 | 557072000 | 6,82161E+13 | 2,34477E+36 | 2,20323E-10 | 6,88508E-15 | 0 | 4161320267 | 2,30224E+11 |
| EV Lacertae | 1,57982E+17 | N/A | N/A | N/A | N/A | N/A | N/A | N/A | N/A | N/A |
| 70 Ophiuchi A | 1,57414E+17 | 1,82988E+30 | 619742600 | 7,06627E+13 | 2,80155E+36 | 2,51357E-10 | 7,8549E-15 | 0 | 4256101563 | 2,62653E+11 |

**Table 1. BER and channel capacity for the 50 stars nearest to the SUN**

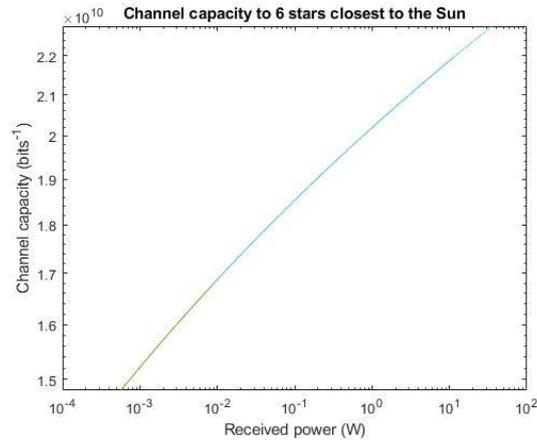

Figure 11. Channel capacity of the six stars nearest to the Sun

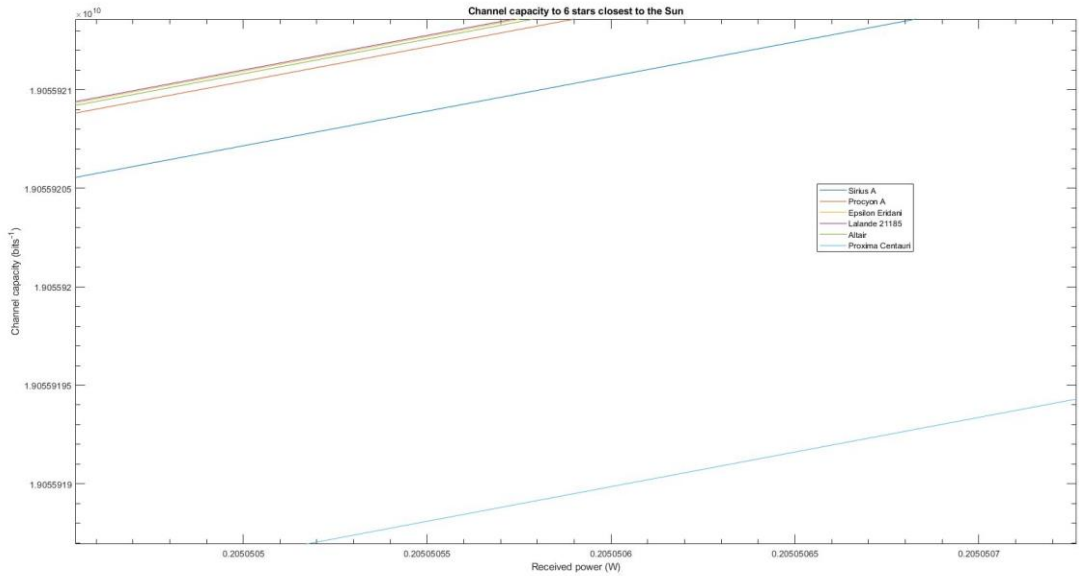

Figure 12. Detail of the channel capacity of the six stars nearest to the Sun